\documentclass[aps,prx,reprint,groupedaddress,dblfloatfix,floatfix]{revtex4-2}

\usepackage{amsmath,amssymb,amsthm}
\usepackage{amsfonts}
\usepackage{bm}
\usepackage{braket}
\usepackage{dsfont} 

\usepackage{graphicx}
\usepackage{float}
\usepackage{placeins}

\usepackage{enumitem}
\usepackage{soul}
\usepackage[normalem]{ulem}
\usepackage{xcolor}
\usepackage{listings}

\usepackage{microtype}

\usepackage[
    colorlinks=true,
    citecolor=blue,
    linkcolor=blue,
    urlcolor=blue
]{hyperref}

\begin{document}

\title{Trotterisation Thresholds and Dissipative Quantum Chaos in Open-System Digital Quantum Simulation}

\author{A.~Manatuly$^{1,2}$}
\affiliation{$^1$Centre for Quantum Software \& Information, Mathematical \& Physical Sciences, University of Technology Sydney, Sydney, Australia. 
$^2$Sydney Quantum Academy, Sydney, Australia.}

\author{L.~M.~Sieberer$^{3,4}$}
\affiliation{$^3$Institute for Theoretical Physics, University of Innsbruck, 6020 Innsbruck, Austria. $^4$PlanQC GmbH, 85748 Garching, Germany.}

\author{C.~Kargi$^{1,2,\dag}$}
\affiliation{$^1$Centre for Quantum Software \& Information, School of Mathematical \& Physical Sciences, University of Technology Sydney, Sydney, Australia. $^2$Sydney Quantum Academy, Sydney, Australia. ($^\dag$Current affiliation: Quantinuum, Cambridge, United Kingdom.)}

\author{N.~K.~Langford$^{1,}$}
\email[Corresponding author: ]{nathan.langford@uts.edu.au}
\affiliation{$^1$Centre for Quantum Software \& Information, School of Mathematical \& Physical Sciences, University of Technology Sydney, Sydney, Australia}

\date{\today}

\begin{abstract}
Digital quantum simulation (DQS) is one of the most promising applications of quantum computing.
Trotterised DQS models quantum dynamics by decomposing the evolution into a sequence of gate-based Trotter steps.
Using larger Trotter steps reduces the required gate count, but it is now understood that the Trotterisation step size cannot be increased beyond a certain threshold value, beyond which there is a sudden breakdown in simulation accuracy, due to the onset of quantum chaos.
To date, this behaviour has only been studied for closed systems not subjected to the environmental noise and decoherence to which real-world quantum computers are highly sensitive.
In this work, we study a range of experimentally relevant open-system quantum simulator models and show that Trotterisation thresholds persist even when open-system effects are included, even in the long-time, steady-state limit, once all coherence has been lost.
We show that open DQS can accurately reproduce the dynamics of open target systems prior to the threshold, but still exhibits a sudden performance breakdown beyond it.
We introduce new, statistical techniques to rigorously characterise the onset of quantum chaos beyond the Trotterisation threshold for both approximately unitary and genuine dissipative quantum chaos, based on random matrix theory (RMT), building on techniques developed for closed systems.
Finally, we demonstrate that dissipative quantum chaos emerges when the dissipation is in balance with the presence of underlying unitary quantum chaos; the dissipation must be strong enough to induce genuinely dissipative behaviour, yet not so strong as to suppress the underlying chaotic structure.
\end{abstract}

\maketitle


\section{Introduction}


Digital quantum simulation (DQS) is a powerful application of quantum computers, enabling the simulation of physical phenomena that are classically intractable~\cite{feynman2018simulating,lloydUniversalQuantumSimulators1996,PhysRevLett.79.2586,georgescu2014quantum,daley2022practical}.
The Trotter-Suzuki decomposition approximates the continuous time-evolution of a quantum system by discretising it into a sequence of quantum gates, each corresponding to evolution over a finite time interval known as the Trotter step size~\cite{trotter1959product,suzuki1990fractal}.
It is one of the most widely used approaches to DQS as it provides leading performance compared to other methods~\cite{childsTheoryTrotterError2021}.
The Trotter step size is a key parameter in optimising DQS protocols, as it is a crucial factor in determining the accuracy of the simulation and the required computational resources.
As real world systems are invariably affected by environmental noise, smaller step sizes also lead to increased errors and ultimately higher resource requirements, as more gates are required to simulate the same total evolution time.
On the other hand, it has also been shown that, at large step sizes, DQS systems experience a breakdown in the accuracy of the simulation due to the onset of quantum chaos~\cite{HeylM2019QLTE,SiebererLM2019DQC,kargi2025quantum}.

Although the Trotterisation threshold has been studied in detail in closed-system quantum simulators~\cite{HeylM2019QLTE,SiebererLM2019DQC,kargi2025quantum}, its behaviour in open-system settings remains largely unexplored.
Real-world digital quantum simulators are inevitably affected by environmental noise and decoherence, which can significantly influence the performance of DQS protocols~\cite{breuer2002theory,RevModPhys.93.025005}.
It is therefore important to determine whether the Trotterisation threshold survives in the presence of environmental noise and decoherence, and how these effects influence the onset of quantum chaos in open-system DQS.
The Trotterisation threshold in closed-system DQS has been studied using various quantities that are experimentally accessible, such as the time evolution of local observables~\cite{HeylM2019QLTE,SiebererLM2019DQC,kargi2025quantum}, out-of-time-ordered correlators~\cite{HeylM2019QLTE,SiebererLM2019DQC}, fidelity~\cite{SiebererLM2019DQC,kargi2025quantum}, and inverse participation ratio~\cite{HeylM2019QLTE,SiebererLM2019DQC,kargi2025quantum}. 
Digital quantum simulation has also been studied experimentally in the presence of hardware noise and decoherence~\cite{kuper2022trottererrorsemergencechaos}.
However, no systematic study has been conducted to characterise the Trotterisation threshold in open-system DQS, and little is known about how the threshold is affected by environmental noise and decoherence.
Furthermore, dissipation does not necessarily only arise in the context of noise and decoherence that is detrimental to performance in digital quantum simulation.
Engineered dissipative processes can also be exploited to prepare and stabilise non-trivial quantum states, including entangled and squeezed states, as well as many-body steady states~\cite{poyatos1996quantum,PhysRevA.61.022301,diehl2008quantum,PhysRevA.78.042307,PhysRevLett.100.140501,verstraete2009quantum,barreiro2011open,PhysRevLett.107.080503,harrington2022engineered}.
It is therefore also highly pertinent to study Trotterisation threshold and performance behaviours in open-system DQS where dissipation is being used as a resource for quantum processing.

In this work, we show that the Trotterisation threshold persists under dissipative dynamics across a broad range of dissipation rates, disappearing only when dissipation becomes so strong as to completely dominate the system dynamics, even on the time-scale of a single Trotter step.
Remarkably, the threshold survives even in the steady state regime, making it experimentally discoverable even through steady-state expectation values of local observables.
We further demonstrate that, in the presence of environmental noise, digital quantum simulators can faithfully reproduce open-system target dynamics, providing a potential route to the preparation and stabilisation of useful states for quantum information processing.
We introduce new, quantitatively rigorous statistical measures of quantum chaos based on random matrix theory (RMT) that are applicable in both approximately unitary and genuinely dissipative regimes.
Using these measures, we show that the threshold for residual unitary quantum chaos coincides with the conventional unitary quantum chaos threshold.
Furthermore, we find that genuinely dissipative quantum chaos also emerges beyond a threshold located at the same Trotter step size.
Specifically, dissipative quantum chaos arises when the dissipation is strong enough to induce genuinely dissipative behaviour, but not so strong as to completely suppress the underlying chaotic structure of the unitary dynamics.


\section{Models and Open-System Dynamics}


\subsection{Open-System Framework}

We consider open quantum systems subject to Markovian dissipation, a widely relevant and applicable model of environmental noise in quantum systems~\cite{breuer2002theory}.
Theoretically, this approximation is generally valid when the system-environment coupling is weak and the environment has a correlation time that is short compared with the characteristic timescales of the system dynamics, conditions that are also almost always true in practice~\cite{breuer2002theory}.
Under these conditions, the system dynamics can be accurately modelled by a Lindblad master equation~\cite{RevModPhys.93.025005,mi2024stable}:
\begin{equation}
	\dot{\rho} = \mathcal{L}\rho = -i\left[H, \rho\right] + \sum_{k} \gamma_k \mathcal{D}(O_k)\rho,
\end{equation}
where $\rho$ is the density matrix, $H$ is the Hamiltonian generating the coherent part of the evolution, $O_k$ are the jump operators associated with each dissipation channel, and $\mathcal{D}(O)\rho := O\rho O^{\dagger} - \frac{1}{2}\left\{O^{\dagger}O, \rho\right\}$ denotes the dissipator, with $\gamma_k$ the corresponding decay rate.
Each gate in the Trotterised evolution is modelled by a propagator generated by the corresponding master equation.
The resulting Floquet superoperator $\mathcal{E}_F$ can therefore be written as
\begin{equation}\label{eq:FloquetSuperoperator}
    \mathcal{E}_F  = \prod_{i}e^{\mathcal{L}_i\tau} \approx e^{\sum_{i}\mathcal{L}_i\tau}, 
\end{equation}
where $\mathcal{L}_i$ describes the open-system evolution for the $i^\text{th}$ gate.
The Floquet superoperator $\mathcal{E}_F$ is not, in general, guaranteed to admit an equivalent description in terms of a Lindblad master equation of the form $\mathcal{E}_F(\rho) = e^{\mathcal{L}_\text{eff}\tau}(\rho)$, where $\mathcal{L}_\text{eff}$ is an effective Lindbladian.
This is because, while the product of completely positive trace-preserving (CPTP) maps is itself a CPTP map, it may not correspond to the exponential of a Lindbladian~\cite{PhysRevB.101.100301,PhysRevB.104.165414}.
As we will see, however, in many cases of practical interest the Trotterised evolution generated by $\mathcal{E}_F$ provides an accurate approximation to the target dynamics, particularly when the Trotter step size $\tau$ is small, as evidenced by the high fidelity between the Trotterised and target evolutions shown later in this work.
This is relevant to the context of simulating open-system target models.
However, regardless of whether an effective Lindbladian exists, the spectral properties of the Floquet superoperator $\mathcal{E}_F$ can be analysed directly to investigate signatures of quantum chaos in the open-system dynamics~\cite{HaakeF2018QSC,braun2000dissipative,PhysRevLett.123.254101,sa2020complex}.

\subsection{Physical Models}\label{Sec:PhysModels}

To develop a comprehensive understanding of digital quantum simulation and investigate Trotterisation thresholds and quantum chaos, we consider three paradigmatic models with diverse structural features, including different symmetries, interaction ranges, and Hilbert space dimensions: (i) the quantum kicked top, (ii) the nearest-neighbour Heisenberg spin chain, and (iii) the Rabi model.
These are the same models previously employed to study closed-system DQS in Ref.~\cite{kargi2025quantum}.
Here, we extend each model to the open-system setting by introducing physically motivated dissipative channels, while retaining the same closed-system Hamiltonians and Trotterisation protocols as in our previous work.
The subscript ``dig'' and ``ide'' refers to the digitised (Trotterised) and ideal (non-digitised) target models, respectively, and we use $\psi$ or $\rho$ to indicate whether the subscripts refer to the closed- or open-system models.
A detailed description of the models and their open-system extensions is provided in the Appendix~\ref{Supp:PhysicalModels}.

Note, in this paper, we define for each of these models a set of reference dissipation rates, chosen for relevance to current circuit QED implementations.
Our results are not restricted to these reference dissipation rates, however:
they serve only to identify values representative of the general behaviour of open-system quantum chaos under physically relevant noise conditions.
Throughout the paper, we then vary overall dissipation levels by multiplying all reference dissipation rates by a scaling factor according to $\gamma_k\rightarrow f_\gamma \gamma_k$, where $\gamma_k$ is the decay rate of the $k^\text{th}$ channel.
The qualitative features identified in our analysis persist across a range of dissipation strengths, provided that the system remains in the weak-coupling regime where the Markovian approximation is valid.


\section{Persistence of the Trotterisation Threshold in Open-System Digital Quantum Simulation}



\begin{figure*}[!t]
	\centering
	\includegraphics{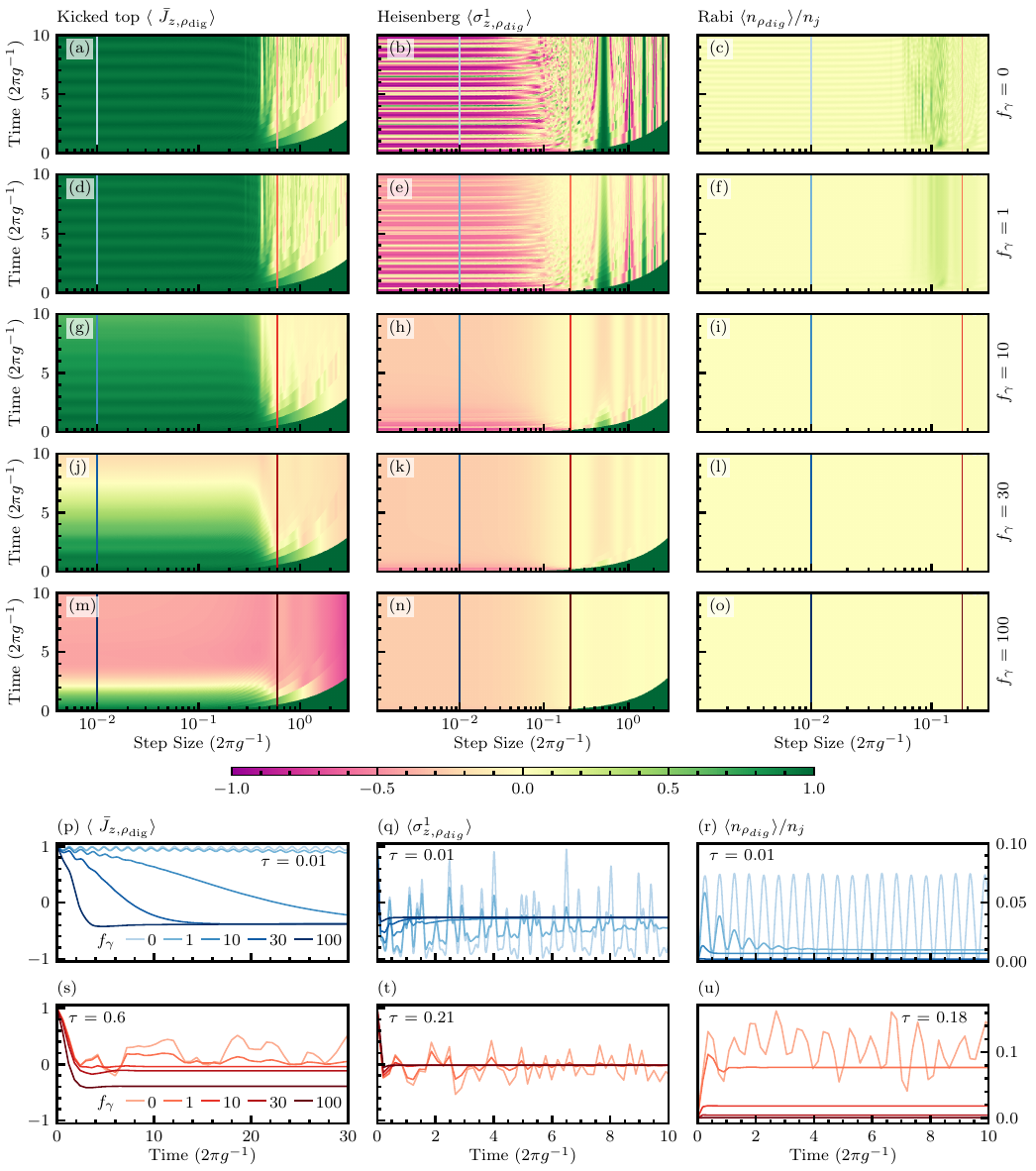}
  \caption[Time-evolution of local observables in open-system DQS]{
    Time-evolution of local observable expectation values in open-system DQS for the kicked top (with $j=10$), Heisenberg (chain of $N=5$ qubits), and Rabi (with cavity truncation dimension $\text{dim}_c=20$) models, corresponding to the first, second, and third columns, with the corresponding local observables magnetisation $\langle\bar{J}_{z}\rangle = \frac{1}{j} \langle J_z \rangle$, polarisation of the first qubit $\langle\sigma_{z}^{1}\rangle$, and normalised photon number $\langle n\rangle/n_j$, respectively.
    Initialisation: in the kicked top model, the spin is initially in a maximal spin state; in the Heisenberg model, only the first qubit in the chain is excited, and the rest are in ground state; in the Rabi model, the qubit is initially excited with zero cavity photons.
    The first five rows show the time evolution of the observables under open-system DQS at different dissipation strength scaling factors $f_\gamma = 0.01, 1, 10, 30, 100$ respectively, plotted as a function of the Trotter step size $\tau$ (x-axes) and simulation time (y-axes).
    The time traces of the observables for different dissipation strengths (p-r) at pre-threshold and (s-u) post-threshold Trotter step sizes. 
  }
	\label{fig:Dynamics}
\end{figure*}


\begin{figure*}[!t]
    \centering

    \includegraphics{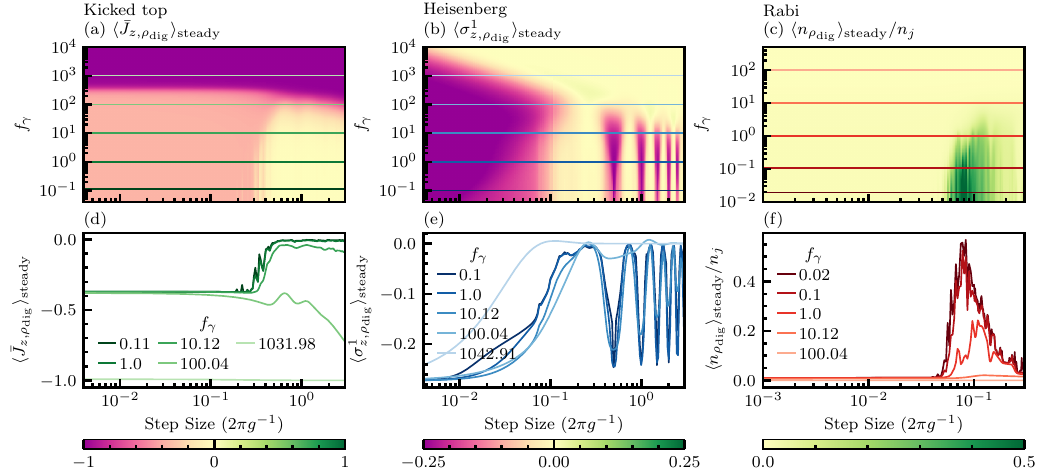}

    \caption[Steady-state observables]{
    Steady-state values of local observable expectation values in open-system DQS for the Kicked top, Heisenberg, and Rabi models with the same parameters as in Fig.~\ref{fig:Dynamics}, corresponding to the first, second, and third columns, with the corresponding local observables magnetisation $\langle\bar{J}_{z}\rangle=\frac{1}{j}\langle J_z\rangle$, polarisation of the first qubit $\langle\sigma_z^1\rangle$, and normalised photon number $\langle n\rangle/n_j$, respectively. Panels (a--c) show the steady-state values as a function of the Trotter step size $\tau$ and dissipation rate scaling factor $f_\gamma$, while (d--f) show traces at selected values of $f_\gamma$.}
    \label{fig:SteadyCombined1}
\end{figure*}


To determine whether the Trotterisation threshold persists in open-system DQS and how it depends on the dissipation strength, we first examine experimentally accessible dynamical and steady-state observables.
Results for our three representative models—the kicked top, Heisenberg chain, and Rabi model—are shown in the first, second, and third columns of Fig.~\ref{fig:Dynamics}, respectively.
The example observables are the spin magnetisation $\langle\bar{J}_{z}\rangle=\frac{1}{j}\langle J_{z}\rangle$ for the kicked top, the polarisation of the first qubit $\langle\sigma_z^1\rangle$ for the Heisenberg chain, and the normalised photon number $\langle n\rangle/n_j$ (where $n_j=7\times\dim_j$) for the Rabi model, where $\dim_j=2j+1=2$ denotes the Hilbert-space dimension of a spin-$\frac{1}{2}$ qubit [see Ref.~\cite{kargi2025quantum} for more details about the photon number normalisation].
The first five rows of Fig.~\ref{fig:Dynamics} show the evolution of these observables under open-system DQS for different dissipation strengths as functions of the Trotter step size $\tau$ (horizontal axis) and simulation time (vertical axis) chosen to illustrate different dynamical regimes.
The final two rows present representative time traces in the pre- and post-threshold regimes.
Similar results for simulation fidelity are given in Appendix~\ref{Supp:FidelityDynamics}.

\subsection{Universal Behaviours}

Three distinct regimes remain visible in the time evolution of the observables even in the presence of dissipation, as shown in Fig.~\ref{fig:Dynamics}.
When the Trotter step size is small, the Trotterised evolution exhibits periodic or quasi-periodic revivals, which decay faster as the dissipation strength is increased; this behaviour is also visible in the time traces of the observables at pre-threshold Trotter step sizes, shown in Fig.~\ref{fig:Dynamics}~(p-r).
This quasi-periodicity indicates that the Trotterised evolution closely approximates the exact dynamics [see Fig.~\ref{fig:Fidelity}].
As the Trotter step size is increased, the period of the quasi-periodic revivals alters, and the revivals become less pronounced, as seen in the corresponding pre-threshold time traces in Fig.~\ref{fig:Dynamics}~(a-f).
The simulation fidelity also shows transition to lower values at these step sizes [see Fig.~\ref{fig:Fidelity}].
After a certain point, this quasi-periodicity is completely lost and the time evolution of the observables becomes irregular, as seen in the corresponding post-threshold time traces in Fig.~\ref{fig:Dynamics}~(s-u).
This loss of regularity signals a breakdown of the Trotterised approximation previously shown to stem from an onset of quantum chaos~\cite{kargi2025quantum}, and is accompanied by a sharp drop in the simulation fidelity [see Fig.~\ref{fig:Fidelity}]. 
The transition between these regimes is sharp enough to be identified by visual inspection alone, indicating the presence of a well-defined Trotterisation threshold in open-system DQS.

Visual comparison between the rows in Fig.~\ref{fig:Dynamics} shows that the position of the Trotterisation threshold does not change obviously with the dissipation strength and matches the position of the threshold in closed-system DQS.
This indicates that the position of the Trotterisation threshold is determined by the underlying unitary dynamics, and is not significantly affected by the presence of dissipation.
However, as the dissipation strength becomes stronger, the Trotterisation threshold becomes less pronounced or even vanishes in the presence of strong dissipation as can be seen in Fig.~\ref{fig:Dynamics}~(m-o). 
This occurs earlier for the Rabi model, due to the amplified impact of dissipation rates in the cavity.
The persistence of the Trotterisation threshold at the same position for different dissipation strengths illustrates that it should be relatively straightforward to observe and locate it experimentally even in the presence of substantial dissipation, as long as it is not too strong.

The fact that the Trotterisation threshold persists in open-system DQS well beyond the timescale over which coherent dynamics decay [Fig.~\ref{fig:Dynamics}] suggests that it may be interesting to study the threshold in the long-time limit.
In fact, looking at the system steady-state regime in Fig.~\ref{fig:SteadyCombined1}, all three models still show a sharp transition in the steady-state values of the observables, provided the dissipation rate scaling factor is not too large. 
When the dissipation is sufficiently weak, all three models exhibit the same qualitative behaviour: over a broad range of dissipation strengths, the Trotterisation threshold remains nearly unchanged; beyond a critical dissipation strength, the threshold shifts to larger Trotter step sizes, before eventually disappearing once dissipation becomes strong enough to dominate the dynamics; in the post-threshold regime, the larger the Trotter step size, the smaller the dissipation strength required for the post-threshold patterns to vanish.
The post-threshold regime also exhibits signatures of stable island behaviours (manifesting as vertical stripes in local observables and fidelity), identified previously in Ref.~\cite{kargi2025quantum}.

The consistent behaviour observed across all three models indicates that these post-threshold features are generic to open-system DQS, rather than being specific to a particular model or dissipation channel.
Observing the threshold in the steady-state regime also shows that coherence is not required to observe the Trotterisation threshold even though the threshold is determined by the underlying unitary dynamics.
This demonstrates that the visibility of the threshold is a matter of the balance between the unitary and dissipative parts of the dynamics, rather than the presence of coherence in the system.

The presence of $\pi/2$ rotation gates in a Trotterised evolution can lead to a change in the orientation of the dissipation operators, which affects the steady-state properties of the system.
At very large dissipation rates, the Trotterisation threshold vanishes for the kicked top and Rabi models, as shown in Fig.~\ref{fig:SteadyCombined1}~(a) and (c).
However, the Heisenberg model exhibits qualitatively different behaviour: the threshold persists even at very large dissipation rates, although it shifts towards smaller Trotter step sizes, as shown in Fig.~\ref{fig:SteadyCombined1}~(b) and (e).
This difference arises because the Trotterised Heisenberg evolution contains rotation gates that effectively change the orientation of the dissipation operators during the evolution. In contrast to the kicked top and Rabi models, where sufficiently strong dissipation drives the system towards a simple ground state, the Heisenberg model retains a non-trivial steady state even at very large dissipation rates.
This steady state is determined by the competition between dissipation channels acting along different axes throughout the Trotterised evolution.


\begin{figure*}[!t]
	\centering
	\includegraphics{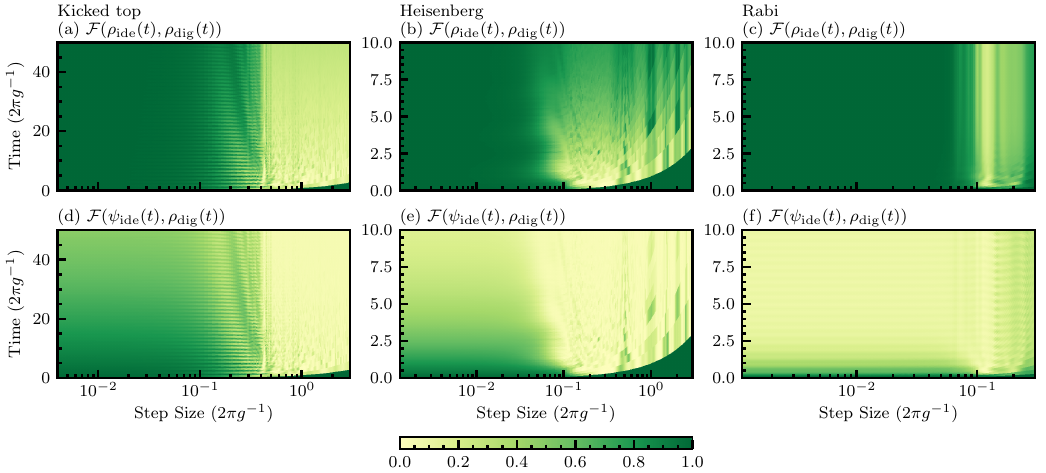}
    \caption[Comparison of simulation fidelities]{Comparison of simulation fidelities of the open-system DQS with respect to the open-system target dynamics and the closed-system target dynamics.
    The subscripts $\mathrm{ide}$ and $\mathrm{dig}$ denote the ideal and digital evolutions, respectively. 
    Panels (a--c) show the simulation fidelity $\mathcal{F}(\rho_{\mathrm{ide}}(t),\rho_{\mathrm{dig}}(t))$ between the open-system target dynamics and the open-system DQS, while panels (d--f) show the simulation fidelity $\mathcal{F}(\psi_{\mathrm{ide}}(t),\rho_{\mathrm{dig}}(t))$ between the closed-system target dynamics and the open-system DQS. 
    The columns correspond, from left to right, to the kicked top, Heisenberg, and Rabi models with the same parameters as in Fig.~\ref{fig:Dynamics}.
    For all open system models, the dissipation rate scaling factor is $f_\gamma = 1$.}

	\label{fig:FideComparison}
\end{figure*}


\subsection{Noisy Unitary DQS vs Open Target Models: Two Open DQS Paradigms}

There are two complementary ways of viewing the role of dissipation in DQS. 
In the first, the target dynamics are those of a closed system, and dissipation constitutes an unwanted source of noise that reduces the fidelity of the simulation. 
In the second, the target dynamics are themselves open-system dynamics, with dissipation forming an intrinsic part of the model. 
A prominent example is engineered dissipation, where dissipative processes are exploited as a resource to prepare and stabilise non-trivial quantum states. 
Although the dissipation described in the models in Appendix~\ref{Supp:PhysicalModels} is of a form typically associated with unwanted experimental noise, it is instructive to also consider the implications of our results in the context of the second paradigm, and compare our digitised open-system models with non-digitized open-system targets. 
Taking this perspective, the results in Appendix Figures~\ref{fig:Fidelity} and~\ref{fig:SteadyCombined2}---which illustrate high fidelities between digitised dynamics and open-system target models---illustrate that DQS platforms can accurately reproduce non-digitised open-system dynamics.

Figure~\ref{fig:FideComparison} shows the illustrative example comparing the fidelity of the open-system DQS with respect to the open- and closed-system target dynamics.
The presence of dissipation unsurprisingly causes rapid decay in fidelity relative to a unitary target model, which simply reflects the usual challenges associated with simulating closed-system dynamics using a real-world quantum processor affected by unavoidable decoherence.  
By contrast, and consistent across all models studied here, fidelities remained high for long times relative to an effective Lindblad open target system, despite there being no guarantee that open Floquet systems have an effective Floquet Lindbladian~\cite{PhysRevB.101.100301,PhysRevB.104.165414}.  
When appropriately designed, open-system DQS can not only accurately simulate open-system target dynamics but also prepare and stabilise non-trivial quantum states, including entangled and squeezed states, as well as many-body steady states~\cite{poyatos1996quantum,PhysRevA.61.022301,diehl2008quantum,PhysRevA.78.042307,PhysRevLett.100.140501,verstraete2009quantum,barreiro2011open,PhysRevLett.107.080503,harrington2022engineered}.   
Such approaches can require correlated dissipative processes that may be challenging to implement, and may not map readily onto natural dissipation channels, but they can reduce or eliminate constraints associated with the coherent elements involved, in exchange for short-time coherence requirements.  
Nevertheless, these results illustrate that real-world decoherence may not always present an insuperable barrier to the realisation of high-fidelity DQS for complex systems, and though there are obviously challenges associated with implementing dissipation in a controllable way, digitisation does offer ways to achieve this.  
Indications of this kind of correspondence have previously been observed in experimental DQS implementations~\cite{LangfordNK2017DQRS}.  
Finally, these results also illustrate the significant potential differences that can arise between the two paradigms of open DQS.


\section{Quantum Chaos in Open-System Digital Quantum Simulation}


In the previous section, we identified the Trotterisation threshold in open-system DQS by analysing the time evolution of local observables and the steady-state properties of the system.
In closed-system DQS, the breakdown of the Trotterised evolution beyond the threshold has been identified as resulting from the onset of quantum chaos~\cite{HeylM2019QLTE,SiebererLM2019DQC,kargi2025quantum}.
In our previous work~\cite{kargi2025quantum}, first indications of that quantum chaos were observed through the destruction of quasi-periodic revivals in the time evolution of local observables, accompanied by a sudden transition in the full dynamics of simulation fidelity, and our new results show the same is also observable in open-system DQS.
As in the closed-system case, however, while indicative signatures of quantum chaos like these are useful for identifying the Trotterisation threshold, they are not sufficient to rigorously characterise the onset of quantum chaos in open-system DQS.

Taking an analogous approach to our previous work~\cite{kargi2025quantum}, a more rigorous characterisation of quantum chaos can be obtained by analysing the statistical properties of the eigenvalues and eigenvectors of the system's evolution operator, which can be compared with the predictions of RMT~\cite{HaakeF2018QSC}.
The main advantage of RMT as a diagnostic of quantum chaos is that it provides a universal framework in which the microscopic details of the system are less important than its underlying symmetries, such as time-reversal symmetry, which determine the corresponding random-matrix universality class (e.g., COE, CUE, or CSE).
In the closed-system case, the evolution operator is unitary, and its eigenvalues lie on the unit circle in the complex plane.
Both level-spacing~\cite{BohigasO1984QCS} and eigenvector statistics~\cite{HaakeF1990RMEF} of unitary operators have been used to identify the onset of quantum chaos in closed-system DQS by comparing them to the predictions of Gaussian or Circular random matrix theory ensembles~\cite{kargi2025quantum, Manatuly2026}.
The open-system evolution operator, on the other hand, is a superoperator with its eigenvalues lying inside the unit circle in the complex plane, and eigenvectors forming a bi-orthogonal set~\cite{HaakeF2018QSC}.
Here, we show how the statistical properties of the complex eigenvalues and eigenvectors of the open-system superoperators can be analysed rigorously and quantitatively to identify the presence of quantum chaos, by comparing them with the RMT predictions of non-Hermitian Ginibre ensembles~\cite{ginibre1965statistical,PhysRevLett.61.1899,PhysRevLett.62.2893,PPeplowski_1993, sa2020complex,PhysRevX.13.031019, PRXQuantum.4.030328}.

\subsection{Quantum Chaos in ``Noisy'' Unitary DQS}

Although open-system evolution is inherently non-unitary, DQS traditionally and most commonly aims to simulate Hamiltonian dynamics while minimising unwanted dissipative effects.
In this context, it is therefore still valuable to study closed-systems behaviours of quantum chaos and to understand how they survive in and are affected by the presence of dissipation.
Challengingly, however, the rigorous RMT signatures of quantum chaos we introduced in~\cite{kargi2025quantum} are explicitly defined in terms of the statistical properties of a unitary matrix, which are, at first glance, immediately violated at a fundamental mathematical level once even small amounts of dissipation are introduced.
Here, we introduce a new statistical approach to characterise how the statistical properties of \emph{unitary} quantum chaos survive in an open-system superoperator, motivated by the recognition that successful implementation of the traditional ``approximately unitary'' DQS paradigm requires any dissipation to be small compared with the coherent evolution.
Under an assumption that the unitary part of a Trotterised evolution captures the dominant features of the system's dynamics, the onset of quantum chaos in open-system DQS can be characterised by isolating the main unitary component from the open-system evolution, and applying our established closed-system statistical techniques to that~\cite{kargi2025quantum}.

\subsubsection{Residual Unitary Quantum Chaos}

To isolate the main unitary component of an open-system evolution, we use the minimal Kraus or operator-sum representation of the open-system Floquet superoperator $\mathcal{E}_F(\rho)$, which, for open quantum processes, is analogous to the minimal pure-state decomposition (or spectral decomposition) of a mixed quantum state.
This is a completely positive trace-preserving (CPTP) map~\cite{langford2007encoding,nielsen2010quantum}:
\begin{equation}\label{eq:KrausEq}
    \mathcal{E}_F(\rho) = \sum_{i=1}^{D^2} \frac{d_i}{D} K_i \rho K_i^\dagger,
\end{equation}
where $D$ is the dimension of the Hilbert space, and $K_i$ are the (ordered) canonical Kraus operators ($d_1 > d_2> \dots >d_{D^2}$) satisfying the completeness relation $\sum_i \frac{d_i}{D} K_i^\dagger K_i = I$, which ensures normalisation of the output state.  
When we choose the Kraus operators to be normalised according to $\text{Tr}(K_i^\dagger K_i)=D$, the coefficients $d_i$ obey $\sum_i d_i = D$.
With these definitions, each Kraus operator $K_i$ can be interpreted as a different probabilistic branch of the evolution, with the probability $d_i/D$ determining the weight of that branch in the overall evolution.

When the dissipation is weak and the Trotter step size is not too large, the Kraus operator $K_1$ is expected to dominate the evolution, with its relative probabilistic weight $d_1/D$ being close to unity.
In this regime, $K_1$ is also expected to be approximately unitary, in the sense that $K_1^\dagger K_1 \approx I$.
These two properties provide complementary measures of how close the full open-system evolution is to an effective unitary equivalent.
Conversely, deviation of $d_1/D$ from unity quantifies the degree to which the open evolution is probabilistic in nature (carried by other Kraus components $K_{j\neq 1}$), while the operator error norm $\Vert K_1^\dagger K_1 - I \Vert/D$ quantifies the degree to which the dominant Kraus operator, $K_1$, itself deviates from unitarity. 
When $d_1/D\approx 1$ and $K_1^\dagger K_1\approx I$, the dynamics are therefore well approximated by a single, almost unitary Kraus component, allowing its eigenvalues and eigenvectors to be analysed using established methods for identifying the onset of quantum chaos in closed-system DQS~\cite{kargi2025quantum}.
In this work, we analyse the eigenvector statistics of $K_1$~\cite{IZRAILEVFM1987CSEF,KusM1988UEV,HaakeF1990RMEF} relative to the predictions of unitary RMT for the appropriate universal RMT ensemble.

The agreement between the eigenvector statistics of the dominant Kraus operator $K_1$ and RMT predictions can be quantified using the $X^2_\mathrm{RMT}$ goodness-of-fit test introduced in Ref.~\cite{kargi2025quantum}, and we follow the procedure for binning the eigenvector components and calculating $X^2_\mathrm{RMT}$ as described there.
Since the $X^2_\mathrm{RMT}$ test statistic quantifies statistical consistency with RMT, and since the underlying reduced $\chi^2_{\rm r}$ probability density which governs the expected distribution of $X^2_\mathrm{RMT}$ outcomes is not sharply peaked (the standard deviation $\Delta\chi^2_{\rm r}(k)=\sqrt{2/k}$ scaling only as $1/\sqrt{k}$ relative to a mean of 1 for $k$ degrees of freedom), there is a finite range of values around $X^2_\mathrm{RMT}=1$ that is consistent with RMT.
To identify whether the dynamics produced by $K_1$ are quantum chaotic, we therefore assess whether observed $X^2_\mathrm{RMT}(K_1)$ outcomes consistently fall within the (full width) 99\% confidence region for a reduced chi-squared distribution with the appropriate $k$, over a finite region of random samples.
Finding too many values of $X^2_\mathrm{RMT}(K_1)\gg 1$ (judged relative to the chosen confidence region) conclusively indicates disagreement with unitary RMT, noting that deviations of $K_1$ from unitarity are also expected to result in deviations from RMT statistical predictions.
Consequently, as long as $K_1$ dominates the evolution, $X^2_{\mathrm{RMT}}(K_1)$ provides a meaningful diagnostic of residual unitary quantum chaos. 
Once the full evolution deviates too far from unitarity, $X^2_{\mathrm{RMT}}(K_1)$ cannot be interpreted reliably in isolation~\cite{Manatuly2026}.


\begin{figure*}[t]
	\centering
	\includegraphics{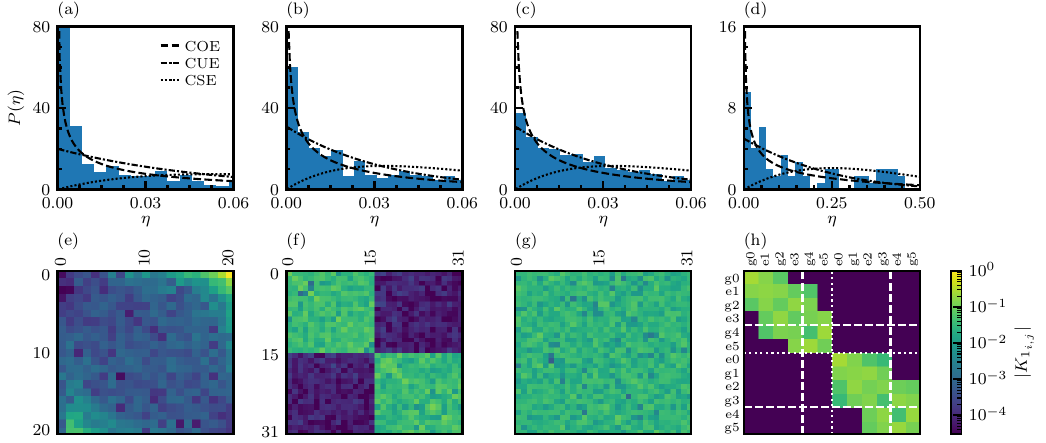}
  \caption[Eigenvector statistics of the dominant Kraus operator]{Example $K_1$ Kraus operators and their eigenvector statistics illustrating the different symmetries which arise in each open-system model.
  (a--d) Eigenvector statistics histograms for the (dominant) $K_1$ Kraus operators compared against COE and CUE distributions, for the Trotterised (a) kicked-top model with $\tau=0.5$ and $f_\gamma=1$; (b, c) Heisenberg model with $\tau=0.2$ and $f_\gamma=1$; and (d) Rabi model with $\tau=0.1$ and $f_\gamma=0.1$.
    (e) Matrix elements of the dominant Kraus operator $K_1$ of the Kicked top model in the computational basis.
    (f),(g) Matrix elements of the dominant Kraus operator $K_1$ of the Heisenberg model in two different bases.
    (h) Matrix elements of the dominant Kraus operator $K_1$ of the Rabi model in the computational basis.
  }

	\label{fig:KrausHist}
\end{figure*}


When comparing system dynamics with RMT, it is important to properly account for any symmetries present,
as they can affect the statistical properties of the dynamics~\cite{dyson1962statistical,HaakeF2018QSC,kargi2025quantum}.
For closed-system dynamics, random Hamiltonians and unitaries exhibit different generic statistical characteristics depending on their RMT universality class, defined by whether they possess an anti-unitary time-reversal symmetry (and which type).
They can also exhibit additional internal structure due to possessing geometric (physical) symmetries, defined under unitary transformations, which constrain dynamical randomness in relation to internal symmetry blocks.
A detailed overview of methods for analysing quantum chaos in closed systems with different symmetries is provided in our previous work~\cite{kargi2025quantum}.

While closed-system symmetries do not necessarily persist in the presence of dissipation, the Kraus operators of an open system's evolution can inherit both statistical characteristics and internal structure from its underlying closed-system dynamics, to the extent that these survive the dissipative perturbation.
This is particularly relevant for the dominant Kraus operator in the weak-dissipation limit relevant to the ``noisy unitary'' paradigm we are studying here.
For example, all the models considered in this work possess time-reversal symmetry in their Hamiltonian limit~\cite{kargi2025quantum}, and we find that an inherited time-reversal symmetry is reflected in the eigenvector statistics of $K_1$, as discussed below.
The Heisenberg and Rabi models also possess unitary parity symmetries in their Hamiltonian limit~\cite{kargi2025quantum}.
When dissipation is introduced, unitary symmetries can be completely broken, or can be either weakly preserved or strongly preserved~\cite{Buca:2012zz}.
An open-system model has a weakly preserved unitary symmetry when the symmetry operator commutes with the full open-system dynamics, including both Hamiltonian terms and dissipators (covariance):
\begin{equation}
   \mathcal{E}_F(S\rho S^\dagger) = S\mathcal{E}_F(\rho)S^\dagger.
\end{equation}
A strongly preserved symmetry requires, in addition to the covariance of a weak symmetry, that all Lindblad jump operators $O_k$ satisfy unitary invariance for the symmetry operator, $[S,O_k]=0$.
For both Heisenberg and Rabi models, parity is preserved as a weak symmetry of their ideal (non-Trotterised) open-system dynamics, but not as a strong symmetry.
For Trotterised open-system dynamics, that weak parity symmetry is then only retained in the Rabi model, but we find the Trotterised Heisenberg model still inherits an approximate weak parity symmetry from its underlying Hamiltonian dynamics.
We discuss below how this is reflected in the structure of the dominant Kraus operator $K_1$.

Figure~\ref{fig:KrausHist} shows example eigenvector component histograms and matrix element norms for the dominant Kraus operator $K_1$ from each Trotterised open-system model at moderate dissipation in the post-threshold regime: (a) at $\tau=0.5$ and $f_\gamma=1$ for the kicked top model; (b, c) at $\tau=0.2$ and $f_\gamma=1$ for the Heisenberg model; and (d) at $\tau=0.1$ and $f_\gamma=0.1$ for the Rabi model.
As shown previously, each of these models exhibits COE symmetry in the Hamiltonian limit~\cite{kargi2025quantum}, and Fig.~\ref{fig:KrausHist}(a), (b) and (d) show that the $K_1$ eigenvector statistics for each model most closely match the COE distributions consistent with the symmetry of their corresponding unitary models.
However, while direct analysis of the kicked-top $K_1$ matrix, shown in Fig.~\ref{fig:KrausHist}(e), is straightforward, additional processing is required for both Heisenberg and Rabi models, due to the impact of their underlying parity symmetries.

In the Trotterised Heisenberg model, the combination of dissipation with $\pi/2$ single-qubit rotations breaks the weak parity symmetry which remains present in the ideal case.
Despite this symmetry being broken in the full Floquet superoperator, however, an examination of the matrix elements of the dominant Kraus operator $K_1$ reveals a residual, approximate block-diagonal structure [Fig.~\ref{fig:KrausHist}~(f)]---where the off-diagonal block components are small, but nonzero---also inherited by its eigenvectors. 
Because only dissipation breaks the parity symmetry, some residual block-diagonality persists in the Kraus operators until dissipation becomes large enough to wash out the underlying parity-conserving unitary dynamics.
If it is not handled properly, this residual parity-conserving structure skews the $K_1$ eigenvector statistics away from the anticipated COE distribution, which creates a barrier to using a $X^2_\mathrm{COE}$ test statistic to detect residual unitary quantum chaos.
If we just ignore the fact that parity symmetry is broken, and simply discard the (small) components from the off-diagonal blocks, Fig.~\ref{fig:KrausHist}~(b) shows that the statistics of the $K_1$ eigenvector components from the diagonal parity-conserving blocks are indeed roughly COE-distributed.
As dissipation increases, however, while $K_1$'s full-matrix statistics still skew away from COE, it becomes difficult to justify ignoring the discrepancy introduced by this Hilbert-space truncation.

In the current context, the primary aim of our RMT analysis of Kraus operators is to identify where any residual unitary quantum chaos (agreement with RMT predictions) is present---not which (if any) symmetry it satisfies---and we can adapt our approach to characterisation accordingly.
In~\cite{kargi2025quantum}, we discuss how analysis of COE statistics in a rotated basis obscures the underlying time-reversal symmetry, and results in eigenvector statistics which match the non-time-reversal symmetric CUE statistics.
Here, we exploit this fact and instead analyse both the Floquet superoperator and the dominant Kraus operator in a rotated basis:
specifically, rotated from the standard (computational) $\sigma_z$ tensor product basis [Fig.~\ref{fig:KrausHist}~(f)], into the (incompatible) rotated $\sigma_x$ product basis [Fig.~\ref{fig:KrausHist}~(g)].
This erases the information available about any residual parity symmetry structure in the system dynamics, eliminating the approximate block-diagonal structure without discarding the information contained in the off-diagonal blocks of the superoperator or Kraus operators.
This does also erase any COE-like symmetry that may still technically be present in the eigenvector statistics, which are now more consistent with the CUE distribution [Fig.~\ref{fig:KrausHist}~(c)]; 
as noted above, this difference is largely irrelevant to the aim of detecting residual unitary quantum chaos.

In contrast, the Trotterised Rabi model does retain its weak parity symmetry, even in the Trotterised dissipative case, so it is still always possible to find a canonical Kraus representation with Kraus operators that respect a definite parity-block structure.
The parity operator $\Pi = e^{i\pi(a^{\dagger}a + \sigma_z + \frac{1}{2})}$ commutes with the qubit dephasing operator ($[\Pi,\sigma_z]=0$).
And while it only \emph{anti-commutes} with both qubit and cavity decay operators, $\Pi a(a^\dagger)\Pi^\dagger=-a(a^\dagger)$ and $\Pi \sigma^{\pm}\Pi^\dagger=-\sigma^{\pm}$, it commutes with their associated dissipators, because the sign changes from the pair of anti-commuting operators in each dissipator term cancel.
This is why the Rabi model parity symmetry is weakly preserved under dissipation (but not strongly), and Fig.~\ref{fig:KrausHist}~(h) illustrates the corresponding, exact parity-block structure in $K_1$'s computational basis matrix elements (indicated by the dotted horizontal and vertical lines).
When analysing the eigenvector statistics of the dominant Kraus operator $K_1$ in the Rabi model, following the method described in~\cite{kargi2025quantum}, we form the eigenvector statistics histogram for $K_1$ from only the components in the non-zero blocks and compare this with the relevant RMT distributions for $D \rightarrow D/2$.
Fig.~\ref{fig:KrausHist}~(d) shows that the Rabi model Kraus operator eigenvector components do still follow the COE distribution, as expected for a system with time-reversal symmetry.
Here, we also use the cavity-specific analysis methods developed for the Rabi model's qubit-cavity system in our previous work, and truncate the Rabi model cavity dimension at $\text{dim}_c=4$ for $X^2_\mathrm{RMT}$ analyses (as explained in~\cite{kargi2025quantum}).

\subsubsection{Analysis of Approximate Unitary DQS}

\begin{figure*}[!t]
    \centering
    \includegraphics{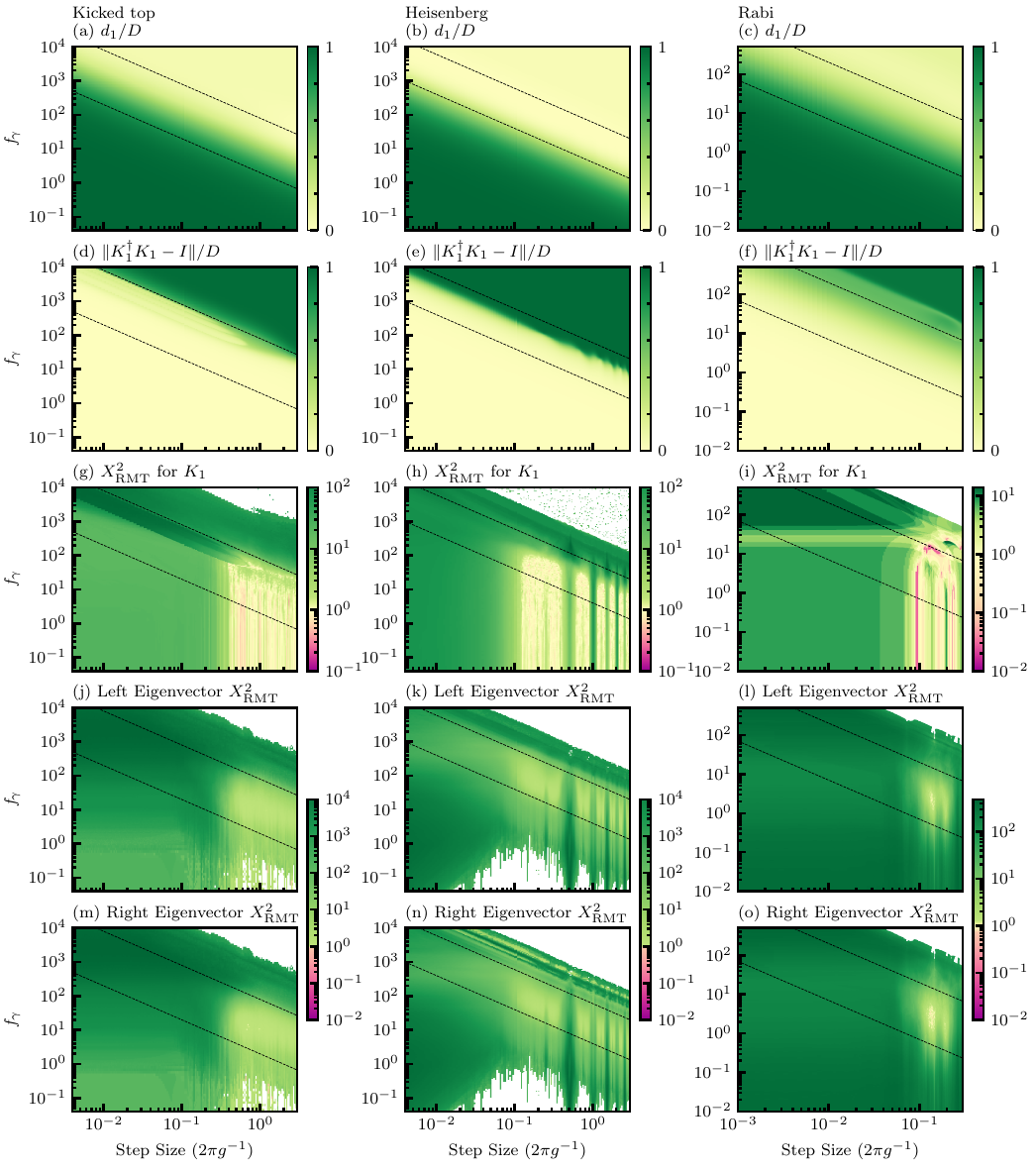}   
    \caption[Residual unitary quantum chaos]{Characterising quantum chaos in open-system DQS. 
    The first, second, and third columns correspond to the kicked top (with $j=10$), Heisenberg (chain of $N=5$ qubits), and Rabi (with cavity truncation dimension $\text{dim}_c=4$) models, respectively.
    The x-axes and y-axes of all panels correspond to the Trotter step size $\tau$ and dissipation strength scaling factor $f_\gamma$, respectively.
    (a-c) Contribution $d_1/D$ of the dominant Kraus operator $K_1$ to the evolution.
    (d-f) Deviation from the unitarity of the dominant Kraus operator $K_1$ quantified using the Frobenius norm as $\Vert K_1^\dagger K_1 - I \Vert / D$.
    (g-i) Goodness-of-fit $X^2_\mathrm{RMT}$ of the eigenvector statistics of the dominant Kraus operator $K_1$ to RMT predictions.
    Goodness-of-fit $X^2_\mathrm{RMT}$ of the (j-l) left and (m-o) right eigenvector statistics of the Floquet superoperator to Ginibre RMT predictions.
    The light-coloured part of the colour bars corresponds to the 99\% confidence regions of the $X^2_\mathrm{RMT}$ test to determine whether the eigenvector statistics are consistent with RMT predictions.
    The dashed lines show the boundaries of the dissipative quantum chaotic regions in the eigenvector statistics of the Floquet superoperator found heuristically from the individual eigenvector statistics of the Floquet superoperator in Fig.~\ref{fig:EigenvectorPanel}.
    }   
    \label{fig:KrausPanel}
\end{figure*}

We now use this technique to analyse our three models for different time discretisation and dissipation rates.
Figure~\ref{fig:KrausPanel}~(a-c) shows the contribution $d_1/D$ of $K_1$ to the evolution versus Trotter step size $\tau$ (x-axes) and dissipation strength scaling factor $f_\gamma$ (y-axes) for the kicked top, Heisenberg, and Rabi models, respectively.
Figures~\ref{fig:KrausPanel}~(d-f) show the corresponding unitarity measure $\Vert K_1^\dagger K_1 - I \Vert/D$.
When the dissipation is weak and the Trotter step size is small, $d_1/D$ approaches unity, indicating that $K_1$ dominates the evolution.
In these regime, $K_1$ is also approximately unitary, as indicated by the close to zero value of $\Vert K_1^\dagger K_1 - I \Vert/D$.
This demonstrates that open-system DQS is well approximated by the dominant Kraus operator in the weak-dissipation, pre-threshold regime, enabling the analysis of residual unitary quantum chaos through the eigenvector statistics of $K_1$.

When the net dissipation per Trotter step is large enough, however, the open DQS superoperator deviates from unitarity in relation to both measures, the intrinsic unitarity of $K_1$ and $K_1$'s probabilistic weight $d_1/D$.
They each decrease with both dissipation strength and Trotter step size, until reaching a saturation value in the dissipation-dominant regime.
These decreases orient relatively consistently along the direction defined by constant ratio $f_\gamma/\tau$.
The contours of constant $d_1/D$ and $\Vert K_1^\dagger K_1 - I \Vert/D$, on the other hand, run approximately diagonally across this direction, parallel to the dashed lines in Figs~\ref{fig:KrausPanel} and~\ref{fig:EigenvectorPanel}, defined by a constant \emph{product} of $f_\gamma\tau$, and selected to align with the boundaries of quantum chaotic regions in the eigenvector statistics of the Floquet superoperator [Fig.~\ref{fig:KrausPanel}~(j-o) and Fig.~\ref{fig:EigenvectorPanel}] (discussed further in Sec.~\ref{subsubsec:analysis-full-open-qchaos}).


In Figure~\ref{fig:KrausPanel}~(g-i), we analyse the eigenvector statistics of $K_1$ using the RMT-based $X^2_\mathrm{RMT}$ goodness-of-fit test with equal-probability bins [See Ref.~\cite{kargi2025quantum} for more details], for the kicked top, the Heisenberg, and the Rabi models, respectively.
The three-colour diverging scale indicates the goodness-of-fit $X^2_\mathrm{RMT}$ between the eigenvector statistics of $K_1$ and RMT predictions, with darker colours indicating stronger disagreement. 
The lighter central colour region corresponds to the 99\% confidence regions of the $X^2_\mathrm{RMT}$ test, within which the statistics are considered consistent with RMT predictions.
A clear threshold can be seen in the eigenvector statistics of the dominant Kraus operator $K_1$ for all three models, consistent with the Trotterisation threshold identified from the dynamical and steady-state observables [Figs.~\ref{fig:Dynamics} and \ref{fig:SteadyCombined1}] and from closed-system DQS.
Before the threshold, the eigenvector statistics exhibit large $X^2_\mathrm{RMT}\gg 1$ values, indicating strong disagreement with RMT predictions.
Beyond the threshold, the statistics become consistent with RMT within the 99\% confidence region, indicating the emergence of quantum chaotic behaviour.

\subsection{Full Open-System Quantum Chaos}

The dominant Kraus operator $K_1$ provides a good approximation of the full evolution in the weak-dissipation regime.
However, as the dissipation strength increases, the contribution of other Kraus operators becomes significant, and the quantum chaotic behaviour of the full open-system evolution can no longer be characterised solely through $K_1$.
Even if individual Kraus components still match RMT predictions for unitary quantum chaos, competition between Kraus components will tend to wash out the coherent effects which distinguish unitary quantum chaos from noise and decoherence.  
As additional Kraus components become significant, however, this could signal a transition to full open-system quantum chaotic dynamics.  
Answering this possibility makes it necessary to investigate the full open Floquet superoperator.

\subsubsection{Statistical Comparison Against Ginibre RMT}

The superoperators of open quantum systems can also be analysed using open-system analogues of the RMT statistics used to characterise unitary quantum chaos in closed quantum systems~\cite{kargi2025quantum}.
Under quantum chaotic conditions, if closed quantum system unitary operators exhibit circular RMT statistics~\cite{HaakeF2018QSC}, then open quantum system superoperators exhibit Ginibre RMT statistics~\cite{ginibre1965statistical,HaakeF2018QSC}. 
Ginibre matrices form Gaussian ensembles of non-Hermitian random matrices with generally complex eigenvalues~\cite{ginibre1965statistical}.
It is important to note that Ginibre matrices do not represent physical generators of dissipative quantum dynamics, as they fail to satisfy the constraints of Hermiticity preservation, trace preservation, and complete positivity required of CPTP superoperators.
Nevertheless, in the open-system quantum-chaotic regime, in the limit of large system size, and under appropriate preselection of the eigenvalues and eigenvectors (discussed below~\cite{Manatuly2026}), it is generally assumed that the overall statistical properties of physical CPTP superoperators converge towards those of the corresponding Ginibre ensembles~\cite{PhysRevLett.61.1899,PhysRevLett.62.2893,PPeplowski_1993,HaakeF2018QSC}.

The key differences of open quantum system superoperators from unitary operators that need to be taken into consideration when analysing their RMT statistics are that their complex eigenvalues are not constrained to lie on the unit circle, and they have both left and right eigenvectors. 
Similar to the closed-system case, both complex level-spacing, and left and right eigenvector statistics, as well as additional eigenvector overlap statistics, can be used to identify quantum chaos by comparing against the corresponding Ginibre RMT statistics~\cite{PhysRevLett.61.1899,PhysRevLett.62.2893,PPeplowski_1993,PhysRevLett.81.3367}.
In this work, we again focus mainly on eigenvector statistics, which provides better statistics due to quadratically more components, but we study both approaches more comprehensively in Ref.~\cite{Manatuly2026}. 
When the system exhibits dissipative quantum chaos, both left and right eigenvector components, $\eta=|c_{in}|$---where \{$c_{in}$\} are the $i$th left or right eigenvector components in the computational basis---are distributed according to~\cite{PPeplowski_1993}:
\begin{equation}
    P(\eta) = 2(N - 1)\eta(1 - \eta^2)^{N-2},
\end{equation}
where $N$ is the dimension, either of the superoperator or of the symmetry subspaces if the system has a unitary symmetry.
Since the Trotterised Rabi model has weak parity symmetry, the Ginibre matrix used for the RMT comparison is half the size of the superoperator, whereas for the Trotterised kicked top and Heisenberg models the Ginibre matrix is the same size as the superoperator.

A superoperator's eigenvectors also need to be properly selected to ensure an apples-to-apples comparison against Ginibre statistics.
Firstly, conservation of the Hermiticity of a density matrix under the evolution of a CPTP map implies that the eigenvalues of the Floquet superoperator are either real, or come in complex conjugate pairs~\cite{HaakeF2018QSC}, and the eigenmatrices of the corresponding complex conjugate eigenvalues are Hermitian conjugates of each other.
To avoid double counting of identical components (and hence introducing correlations that would skew statistical comparisons), only components from one eigenmatrix of each complex-conjugage pair of eigenvalues should be included.
This is implemented most straightforwardly by filtering a superoperator's eigenvectors on whether their eigenvalues have either positive or negative imaginary components.
We then follow standard methods for identifying and analysing only eigenmatrices corresponding to ``bulk'' eigenvalues~\cite{HaakeF2018QSC,PPeplowski_1993}.
A detailed procedure for selecting and analysing eigenvector components, as required for our chi-squared statistical RMT tests, is discussed in~\cite{Manatuly2026}.


\begin{figure*}[!t]
    \centering
    \includegraphics{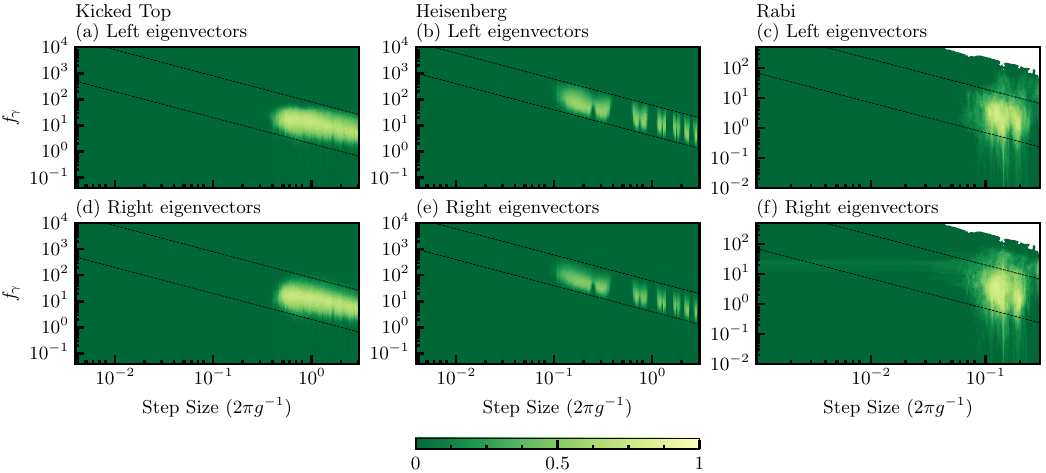}   
    \caption[Random eigenvectors]{Fraction of random eigenvectors of the Floquet superoperator.
    The first, second, and third columns correspond to the kicked top (with $j=10$), Heisenberg (chain of $N=5$ qubits), and Rabi (with cavity truncation dimension $\text{dim}_c=4$) models, respectively.
    The x-axes and y-axes of all panels correspond to the Trotter step size $\tau$ and dissipation strength scaling factor $f_\gamma$, respectively.
    The dashed lines are the constant dissipation lines and highlight the boundaries of the region where the eigenvectors become random.
    }
    \label{fig:EigenvectorPanel}

\end{figure*}


\subsubsection{Analysis of Full Open Quantum Chaos in DQS}
\label{subsubsec:analysis-full-open-qchaos}

We now study the onset of genuine dissipative quantum chaos in open-system DQS using the same DQS models and the eigenvector statistics of the full Floquet superoperator.
Figure~\ref{fig:KrausPanel}~(j-o) shows the left and right eigenvector statistics of the Floquet superoperator of (j),(m) the kicked top model, (k),(n) the Heisenberg model, and (l),(o) the Rabi model as a function of the Trotter step size $\tau$ (x-axes) and dissipation strength scaling factor $f_\gamma$ (y-axes).
Again, the light-coloured part of the colour bar corresponds to the 99\% confidence regions of the $X^2_\mathrm{RMT}$ test to determine whether the eigenvector statistics are consistent with RMT predictions, and the dark regions corresponds to large values of the $X^2_\mathrm{RMT}$ signifying regular dynamics.
Since the number of eigenvector components varies with the parameters, the 99\% confidence region is defined in terms of the number of eigenvector components of the eigenvectors selected for the RMT comparison.
The white regions in the colour maps correspond to values where no valid eigenvectors are identified. 
When both the dissipation rate and the Trotter step size are large, the eigenvalues of the Floquet superoperator collapse to the origin, and when the dissipation is small, the eigenvalues are located near the unit circle and are almost one dimensional.
In both cases, it becomes challenging to identify valid eigenvectors for the RMT comparison, and therefore the corresponding regions are left white in the colour maps.

All the models exhibit a similar pattern: when the Trotter step sizes are small, the $X^2_\mathrm{RMT}$ statistic values are large for all dissipation rate scaling factors.
As the Trotter step size is increased, at sufficiently large dissipation rates, the $X^2_\mathrm{RMT}$ test statistics exhibit drastic changes, showing that all models transition towards close agreement with the predictions of Ginibre RMT, and thus indicating dissipative quantum chaos. 
Importantly, for all models, the Trotter step size where this transition occurs coincides with the threshold of the dominant Kraus operator eigenvector statistics observed in Fig.~\ref{fig:KrausPanel}~(g-i).
The transition in the Heisenberg model is not as sharp as in the other two models, consistent with the fact that fewer eigenvectors become random (as will be discussed further below).

An important feature of the $X^2_\mathrm{RMT}$ results for the left and right eigenvectors is that the dissipative quantum chaotic regions in the eigenvector statistics show a downward trend with increasing Trotter step size. 
The boundaries of these regions are highlighted with dashed lines, provided as ``guides to the eye", showing constant-value contours of the product $f_\gamma\tau$, in locations chosen to align roughly with the boundaries of the region where the eigenvectors become random (again, discussed below).
These results show that, for step sizes beyond the threshold for unitary quantum chaos, the onset of genuine dissipative quantum chaos requires a sufficiently large \emph{total dissipation per Trotter step}, $f_\gamma\tau$, as determined by both dissipation strength and Trotter step size. 
Once the dissipation becomes too strong, however, the $X^2_\mathrm{RMT}$ test statistic again increases to large values, indicating that too much dissipation also prevents the system exhibiting full dissipative quantum chaos, even on the time-scale of a single Trotter step.  
The dashed lines in Figs~\ref{fig:KrausPanel}(g, i) suggest that this occurs the same dissipation rate that the residual unitary quantum chaos in $K_1$ disappears~\cite{Manatuly2026}.

Since the eigenvector statistics of the Floquet superoperator in Fig.~\ref{fig:KrausPanel}~(j-o) are analysed for the selected eigenvectors, it is important to ensure that the number of random eigenvectors is sufficiently large to be representative of the full dynamics of the system.
Fig.~\ref{fig:EigenvectorPanel} shows the fraction of left and right RMT eigenvectors of the Floquet superoperator of (a),(d) the kicked top model, (b),(e) the Heisenberg model, and (c),(f) the Rabi model as a function of the Trotter step size $\tau$ (x-axes) and dissipation strength scaling factor $f_\gamma$ (y-axes).
The fraction of eigenvectors agreeing with Ginibre RMT is defined in terms of the half of the total number of eigenvectors of the Floquet superoperator, since only half the eigenvectors (corresponding to one of each complex-conjugate pair of eigenvalues) are independent, as discussed above.
An individual eigenvector is considered random if its $X^2_\mathrm{RMT}$ statistic value is within the 99\% confidence region of the RMT predictions.
The dashed lines correspond to the constant values of the product $f_\gamma\tau$ and empirically illustrate the boundaries of the region where the eigenvectors become random.

The patterns are similar to those in Fig.~\ref{fig:KrausPanel}~(j-o), and show that, beyond the threshold and at large enough dissipation rates, a significant portion of the eigenvectors of the Floquet superoperator exhibit RMT statistics.
The fraction of random eigenvectors is fewer for the Heisenberg model compared to the other two models, which explains the less sudden transition observed in Fig.~\ref{fig:KrausPanel}~(k,n).
These results demonstrate that studying individual eigenvector statistics of the Floquet superoperators can provide valuable additional information for identifying the dissipative quantum chaotic dynamics in open quantum systems, as contribution of random eigenvectors may be washed out by non-random eigenvectors if the statistics are analysed together in the full collective $X^2_\mathrm{RMT}$. 
For example, on a technical level, the fraction of RMT eigenvectors, particularly for systems like the Heisenberg model, may increase for larger system sizes.
Due to the computational complexity of modelling open-system quantum evolution, the system dimensions considered in this paper are significantly smaller than the sizes we previously analysed for closed systems~\cite{kargi2025quantum}.
More fundamentally, the fact that not all eigenvectors become random may also be connected to the fact that open-system quantum chaos is a more complex phenomenon than unitary quantum chaos, with open-system superoperators having more physical constraints than (and therefore representing strictly only a subset of) the Ginibre matrices used for the RMT comparison~\cite{Manatuly2026}.


\section{Discussion}


In our previous work, we demonstrated the universality of Trotterisation performance and threshold behaviours in closed-system DQS~\cite{kargi2025quantum}, and our results here provide strong evidence that this universality can be extended to open-system DQS.
This shows that the Trotterisation threshold---driven by underlying unitary quantum chaos---is a robust feature of DQS, even in the presence of dissipation. 
This is supported by the fact that similar to the closed-system DQS, the open-system DQS also exhibits periodic or quasi-periodic behaviour in the pre-threshold regime [Fig.~\ref{fig:Dynamics}], where the simulation fidelity is high in both the full dynamics [Fig.~\ref{fig:Fidelity}] and steady state regime [Fig.~\ref{fig:SteadyCombined2}], and a sharp transition to irregular dynamics and low fidelity in the post-threshold regime.
This illustrates that Trotter errors remain \emph{well-behaved} in the pre-threshold regime, while they become large and uncontrollable in the post-threshold regime, even in the presence of dissipation, consistent with the behaviour of Trotter errors in closed-system DQS~\cite{SiebererLM2019DQC,kargi2025quantum}.
Furthermore, the fact that open-system DQS also exhibits signatures of quantum chaos beyond the threshold, in addition to consistent dynamical threshold and pre-threshold behaviours before it, illustrates that the onset of quantum chaos is a universal feature of both open- and closed-system DQS.
And regardless of whether the system is open or closed, our results show that the threshold is driven by the underlying unitary quantum chaos in the system. 

The observation of persistent Trotterisation thresholds in open-system DQS effectively identifies the operating limit for open-system DQS---the usable range of Trotter step sizes that avoids the breakdown of simulation performance---and highlights the necessity for doing so. 
This applies to any DQS algorithms that employ dissipation~\cite{barreiro2011open,PRXQuantum.3.010320}, such as dissipative state preparation~\cite{PhysRevA.61.022301,mi2024stable}, simulation of dynamical maps and quantum channels~\cite{schindler2013quantum}, and simulation of non-equilibrium dynamics~\cite{fauseweh2021digital}.
As demonstrated in~\cite{kargi2025quantum}, the emergence of quantum chaos leads to rapid proliferation of Trotter errors which are not correctable or controllable, as they are at pre-threshold step sizes.
Furthermore, because the open DQS threshold can still be observed even in the steady state regime, and is still coincident with the closed-system threshold, this allows noisy digital quantum simulators to be used to identify fundamental performance limits that will still apply to much higher-fidelity quantum processors.
This could potentially translate into direct resource savings by avoiding carrying out tasks taking up valuable runtime and running costs on expensive large-scale error-corrected, fault-tolerant quantum processors.

The persistence of the Trotterisation threshold in the presence of dissipaton also has other potentially important implications for the practical implementation of DQS, where dissipation is unavoidable.
At the simplest level, it is important to distinguish that the threshold location does not depend on dissipation, because while errors arising from dissipative processes can in principle be suppressed through quantum error correction, Trotter errors are intrinsic to the algorithm and cannot be eliminated in this way.
This is particularly relevant for resource optimisation in DQS, as it could be an important consideration in identifying an optimal Trotter step size to balance the trade-off between simulation accuracy and computational overhead, which increases in the presence of dissipation due to the need for error correction and mitigation techniques~\cite{shor1996fault,preskill1998fault,PhysRevA.57.127,roffe2019quantum}.

In all three models studied in this paper, the emergence of dissipative quantum chaos, demonstrated by $X^2_\mathrm{RMT} \sim 1$ values in the left and right eigenvector statistics [Fig.~\ref{fig:KrausPanel}~(j-o)], occurs at roughly the same Trotter step size as where the eigenvectors of the dominant Kraus operator begin to agree with the RMT distribution [Fig.~\ref{fig:KrausPanel}~(g-i)]. 
This shows that dissipative quantum chaos emerges only if the underlying unitary part of the dynamics is itself quantum chaotic.
The chaos observed, via its eigenvector statistics, beyond the threshold in the open system, corresponds closely to closed-system chaos when the dissipation is weak, because the evolution of the system is well approximated by a single, dominant Kraus operator. 
True dissipative quantum chaos emerges only when the dissipation rate is sufficiently strong, and provided the underlying unitary contributions to the dynamics still exhibit the properties of closed-system quantum chaos.

Our results also highlight the need for further investigation into the interplay between dissipation and quantum chaos in open-system DQS.
On the one hand, our work provides a deeper understanding of the fundamental limits of different paradigms of open-system DQS, and how that relates to both unitary and dissipative quantum chaos.
This could help inform the design of both more robust unitary DQS algorithms, and also more sophisticated DQS algorithms where dissipation is exploited as a beneficial resource~\cite{barreiro2011open,schindler2013quantum,PRXQuantum.3.010320,PhysRevLett.107.080503,harrington2022engineered,mi2024stable}.
But perhaps this can also be pushed further.
For example, in the simulation of complex quantum systems, the presence of dissipation can lead to rich and non-trivial dynamics that are difficult to capture using classical methods.
Does dissipative quantum chaos flag the presence of open-system dynamics that are classically intractable to simulate, or even sample, in the same way that unitary quantum chaos and RMT dynamics have been used to define classically prohibitively complex computational tasks~\cite{BoixoS2018QSNT,Arute2019QS,WuY2021QASC,ZhuQ2021QCA60}?
Could this be directly exploited to provide a novel pathway towards quantum advantage for open-system DQS, even in the presence of dissipation? 
Overall, our results provide new insights that are important not only for the fundamental understanding of quantum chaos in open systems, but could also help develop a better understanding about potential applications of both open quantum chaos and open DQS, and novel opportunities for quantum advantage to be derived from each.


\begin{acknowledgments}
The authors thank T~Srivipat and DW~Berry for useful discussions.

This work has been funded by NKL's Australian Research Council Future Fellowship (FT170100399) and Discovery Project (DP210101367).
AM and CK acknowledge support from UTS President's scholarships, international research scholarships, and Sydney Quantum Academy.
LS acknowledges support from the Austrian Science Fund (FWF) through the project 10.55776/COE1, and from the European Union - NextGenerationEU.
\end{acknowledgments}



\appendix


\section{Physical Models\label{Supp:PhysicalModels}}

\subsection{Quantum Kicked Top Model}

Kicked top models serve as canonical examples of systems that display chaotic dynamics in both the classical and quantum regimes~\cite{HaakeF1987CQCKT,KusM1988UEV,HaakeF2018QSC,SiebererLM2019DQC}.
Quantum-chaotic behaviour emerges in this single-body system~\cite{HaakeF2018QSC,SiebererLM2019DQC}, which has been experimentally realised using a single nuclear spin~\cite{MourikV2018QCNS} and superconducting qubits~\cite{NeillC2016EDQS}.

For the kicked top model, we consider the following Trotterisation:
\begin{equation}
    F(\tau) = e^{-i H \tau} \approx e^{-i H_z \tau} e^{-i H_x \tau},
\end{equation}
where $H_{KT} = H_z + H_x$ is the effective Hamiltonian of the kicked top model, with
\begin{equation}
    H_\mu = \omega_\mu J_\mu + \frac{g_\mu J_\mu^2}{2j+1}, \quad \mu = x,z.
\end{equation}
For clarity, the Trotterisation above is written in its purely unitary form; in the open-system setting considered in this work, each factor $e^{-i H_\mu \tau}$ is instead replaced by a dissipative channel that includes both the unitary evolution generated by $H_\mu$ and associated dissipation, which is discussed further below.
Here, the term proportional to $J_\mu^2$ represents the kick in $H_\mu$, while the term proportional to $J_\mu$ represents a uniform field along the $\mu$-axis.
We use the same simulation parameters as in Refs.~\cite{SiebererLM2019DQC,kargi2025quantum} to facilitate direct comparison with previous results: $g_z := g$, $g_x = 0.7g$, $\omega_z = 0.3g$, and $\omega_x = 0.1g$.
Numerical simulations of open quantum systems are challenging, because the dimension of the associated superoperators scales as the square of the Hilbert space dimension, and hence the number of elements scales as the fourth power.
This scaling makes simulating large systems computationally expensive.
Here, we analyse the system up to a moderate spin size of $j=10$, which is large enough to exhibit quantum chaotic behaviour~\cite{kargi2025quantum}.

Both decay $\mathcal{D}(J_-)$ and dephasing $\mathcal{D}(J_z)$ dissipators are included in the evolution generated by $H_\mu$, with dissipation rates $\gamma_1/(2j)$ and $\gamma_2/(2j)$ respectively.
The operator $J_-$ is the spin-lowering operator, defined as $J_- = J_x - iJ_y$.
The factor $1/(2j)$ ensures that, for $j = 1/2$, the dissipation rates reduce to those of a single qubit.
This allows $\gamma_1$ and $\gamma_2$ to be set to representative values for superconducting transmon devices:
specifically, with a relaxation time of $T_1 = 22\,\mu\mathrm{s}$~\cite{mi2024stable}, and a dephasing time to $T_2 = T_1/1.5 \approx 14.67\,\mu\mathrm{s}$, chosen to be moderately shorter than $T_1$, consistent with typical transmon behaviour~\cite{PhysRevLett.107.240501, mi2024stable}.
To convert these physical timescales into the dimensionless units used in our simulations, we assume a coupling strength of $g = 40\,\mathrm{MHz}$~\cite{mi2024stable}, from which $\gamma_1$ and $\gamma_2$ are obtained relative to $T_1$ and $T_2$, respectively.

For $j \neq 1/2$, the scaling $1/(2j)$ is required to ensure the dissipator terms scale consistently with the unitary part of the dynamics. 
The kicked-top Hamiltonian scales linearly with spin size, $H_{KT} \sim j$, whereas the dissipators, which are quadratic in the jump operators $J_-$ and $J_z$, scale as $j^2$. 
Scaling of dissipation rates inversely with $j$ therefore compensates for this discrepancy, resulting in dissipative terms, $\frac{\gamma}{2j}\mathcal{D}(J)$, that also scale linearly with $j$.

\subsection{Nearest-Neighbour Heisenberg Model}

The Heisenberg model has been widely studied in condensed-matter physics as a model for interacting spins on a lattice~\cite{auerbach2012interacting}.
In this work, we consider a one-dimensional chain of $N$ qubits with open boundary conditions subject to a uniform field and nearest-neighbour interactions, which has already been experimentally implemented using superconducting qubits~\cite{Salathe2015DQS}.
The Hamiltonian of the system is given by:
\begin{equation}
    H_H = \frac{\omega}{2} \sum_{j=1}^N \sigma_z^j + g \sum_{j=1}^{N-1} (\sigma_x^j \sigma_x^{j+1} + \sigma_y^j \sigma_y^{j+1} + \sigma_z^j \sigma_z^{j+1}),
\end{equation}
where $\sigma_\alpha^j$ ($\alpha = x,y,z$) are the Pauli operators acting on the $j$-th qubit, $\omega$ is the uniform field strength, and $g$ is the nearest-neighbour coupling strength.
The total Hamiltonian can be decomposed into $H_H = H_z + H_{xy} + H_{xz} + H_{yz}$~\cite{Heras2014DQS,Salathe2015DQS}, where $H_z = \frac{\omega}{2} \sum_{j=1}^N \sigma_z^j$, $H_{xy} = \frac{g}{2} \sum_{j=1}^{N-1} (\sigma_x^j \sigma_x^{j+1} + \sigma_y^j \sigma_y^{j+1})$, and similarly for $H_{xz}$ and $H_{yz}$.
$H_z$ can be implemented using single-qubit $z$-rotation phase gates, and $H_{xy}$ corresponds to implementing standard two-qubit exchange gates.
$H_{xz}$ and $H_{yz}$ are then realised by applying appropriate single-qubit rotations before and after the two-qubit exchange gates~\cite{Salathe2015DQS}:
\begin{equation}
	H_{xz} = \mathcal{R}_{x,\tfrac{\pi}{2}}^{\forall k}H_{xy}\mathcal{R}_{x,\tfrac{\pi}{2}}^{\forall k \dagger}
	= g^{\prime}\sum_{k=1}^{N-1}\left( \sigma_{x}^{k}\sigma_{x}^{k+1} + \sigma_{z}^{k}\sigma_{z}^{k+1} \right),
\end{equation}
and
\begin{equation}
	H_{yz} = \mathcal{R}_{y,\tfrac{\pi}{2}}^{\forall k}H_{xy}\mathcal{R}_{y,\tfrac{\pi}{2}}^{\forall k \dagger}
	= g^{\prime}\sum_{k=1}^{N-1}\left( \sigma_{y}^{k}\sigma_{y}^{k+1} + \sigma_{z}^{k}\sigma_{z}^{k+1} \right),
\end{equation}
where $g^{\prime} = g/2$, and $\mathcal{R}_{\mu,\theta}^{\forall k} = \prod_{k=1}^{N}e^{-i\frac{\theta}{2} \sigma^{k}_{\mu}}$ (for $\mu\in\{x, y, z\}$) are locally applied, single-qubit rotations by an angle $\theta$ around the $\mu$ axis.

The Trotterised Floquet unitary for one time step is given by:
\begin{equation}
\begin{split}
    F = {} & \prod_{j \,\mathrm{even}} U_{yz}^{j,j+1}(\tau) \prod_{j \,\mathrm{odd}} U_{yz}^{j,j+1}(\tau) \\
    & \times \prod_{j \,\mathrm{even}} U_{xz}^{j,j+1}(\tau) \prod_{j \,\mathrm{odd}} U_{xz}^{j,j+1}(\tau) \\
    & \times \prod_{j \,\mathrm{even}} U_{xy}^{j,j+1}(\tau) \prod_{j \,\mathrm{odd}} U_{xy}^{j,j+1}(\tau)\, e^{-i H_z \tau},
\end{split}
\end{equation}
where $U_{xy}^{j,j+1}(\tau) = e^{-i H_{xy}^{j,j+1} \tau}$, $U_{xz}^{j,j+1}(\tau) = e^{-i H_{xz}^{j,j+1} \tau}$, and $U_{yz}^{j,j+1}(\tau) = e^{-i H_{yz}^{j,j+1} \tau}$ are two-qubit exchange gates acting on qubits $j$ and $j+1$, for a Trotter step size $\tau$. 
Here, the products over even and odd $j$ run over all valid nearest-neighbour bonds $(j, j+1)$ within the chain, so that each pair of neighbouring qubits is coupled exactly once per product.
The reason for this choice of Trotterisation is that the Hamiltonians $H_{ab}^{j,j+1}$, for all $ab \in \{xy, xz, yz\}$, commute with each other when the corresponding indices $j$ are either all even or all odd. 
This allows each coupling type to be activated in only two steps, even and odd, rather than sequentially across all bonds. 
Such parallel activation is important in the presence of dissipation, since it reduces the total duration of each Trotter step compared to sequential activation, hence also the accumulated dissipation.

Each qubit experiences both decay $\mathcal{D}(\sigma_-)$ and dephasing $\frac{1}{2}\mathcal{D}(\sigma_z)$, with dissipation rates $\gamma_1$ and $\gamma_2$ respectively.
We use the same experimentally relevant dissipation timescales and coupling strength as in the kicked top model discussed above: $T_1 = 22\,\mu\mathrm{s}$, $T_2 = T_1/1.5 \approx 14.67\,\mu\mathrm{s}$, and $g = 40\,\mathrm{MHz}$~\cite{mi2024stable}.
In our numerical simulations, both $\omega$ and $g$ are set to $1$ in dimensionless units, with $g = 40\,\mathrm{MHz}$ setting the physical scale from which the numerical values of $\gamma_1$ and $\gamma_2$ are obtained relative to $T_1$ and $T_2$, respectively.
The number of qubits is chosen to be $N=5$, low enough to make the computational complexity manageable, but still large enough that the closed-system unitary exhibits quantum chaotic behaviour~\cite{kargi2025quantum}.

\subsection{Rabi Model}

The Rabi model describes light-matter interaction~\cite{PhysRev.49.324,PhysRev.51.652}, and consists of a single qubit coupled to a single bosonic mode.
This model is fundamental in quantum optics and has been extensively studied in various contexts, including cavity quantum electrodynamics, quantum information processing, and many-body physics~\cite{haroche2006exploring,gardiner2004quantum}. 

The Hamiltonian of the Rabi model is given by:
\begin{equation}
    H_R = \omega_c a^\dagger a + \frac{\omega_q}{2} \sigma_z + g (a + a^\dagger)(\sigma_+ + \sigma_-),
\end{equation}
where $a$ ($a^\dagger$) are the annihilation (creation) operators for the bosonic mode with frequency $\omega_c$, $\sigma_z$ is the Pauli-$z$ operator, $\omega_q$ is the qubit transition frequency, $\sigma_\pm$ are the raising and lowering operators, and $g$ is the coupling strength between the qubit and the bosonic mode.

When the coupling strength $g$ is much smaller than the qubit and bosonic mode frequencies ($g \ll \omega_q, \omega_c$), the rotating-wave approximation (RWA) can be applied, leading to the Jaynes-Cummings (JC) model~\cite{1443594}, which conserves the total excitation number.
The Hamiltonian under the RWA becomes:
\begin{equation}
    H_\mathrm{JC} = \omega_c a^\dagger a + \frac{\omega_q}{2} \sigma_z + g (a \sigma_+ + a^\dagger \sigma_-).
\end{equation}

While physical systems are almost always restricted to the JC regime, the full Rabi model can be reconstructed from the JC model using digital quantum simulation~\cite{MezzacapoA2014DQRS}. 
In particular, the Rabi Hamiltonian can be decomposed into two parts, $H_{R} = H_\mathrm{JC}(g, \Delta_{c}, \Delta_{q}^\mathrm{JC}) + H_\mathrm{AJC}(g, \Delta_{c}, \Delta_{q}^\mathrm{AJC})$, where the anti-Jaynes-Cummings (AJC) Hamiltonian is given by:
\begin{equation}\label{Eq:AJCHamiltonian}
\begin{split}
	H_\mathrm{AJC} &= \mathcal{R}_{x,\pi} H_\mathrm{JC}\mathcal{R}_{x,\pi}^{\dagger} \\
	&= \Delta_{c} a^{\dagger} a - \frac{\Delta_{q}^\mathrm{AJC}}{2} \sigma_z + g(a^{\dagger} \sigma_{+} + a\sigma_{-}),
\end{split}
\end{equation} 
and is obtained by applying a global $\pi$ rotation, $\mathcal{R}_{x,\pi}$, to the qubit~\cite{MezzacapoA2014DQRS}.
Here, $\Delta_{\mu} = \omega_{\mu} - \omega_\mathrm{RF}$ ($\mu = c,q$) represent the instantaneous lab-frame detunings of the bosonic mode and the qubit from the reference frequency $\omega_\mathrm{RF}$ in the rotating frame.
By independently tuning this lab-frame qubit detuning during each simulation phase, we set the specific values for $\Delta_q^\mathrm{JC}$ and $\Delta_q^\mathrm{AJC}$.
If the durations of the JC and AJC evolutions are equal, the effective cavity and qubit frequencies in the Rabi model are given by $\omega_c^\mathrm{eff} = 2\Delta_c$ and $\omega_q^\mathrm{eff} = \Delta_q^\mathrm{JC} - \Delta_q^\mathrm{AJC}$, respectively.
In our simulations we set $\Delta_q^\mathrm{JC} = \Delta_c = 3.5g$ and $\Delta_q^\mathrm{AJC}=0$, consistent with the target model being non-quantum chaotic [see Ref.~\cite{kargi2025quantum}].

The digital quantum simulation of the Rabi model can be performed by Trotterising the evolution under the JC and AJC Hamiltonians~\cite{MezzacapoA2014DQRS}:
\begin{equation}\label{eq:RabiTrot}
\begin{split}
    F(\tau) &= U_\mathrm{JC}(\tau/2) U_\mathrm{AJC}(\tau) U_\mathrm{JC}(\tau/2) = \\
    &= e^{-i H_\mathrm{JC} \tau/2} \mathcal{R}_{x,\pi} e^{-i H_\mathrm{JC} \tau} \mathcal{R}_{x,\pi}^{\dagger} e^{-i H_\mathrm{JC} \tau/2},
\end{split}
\end{equation}
where $\tau$ is the Trotter step size. 
The experimental realisation of the Rabi model using superconducting circuits has been demonstrated in~\cite{LangfordNK2017DQRS}.

Since we are considering open-system DQS, dissipators are added to the evolution generated by the JC Hamiltonian in Eq.~\ref{eq:RabiTrot}. 
The qubit experiences decay $\mathcal{D}(\sigma_-)$ and dephasing $\frac{1}{2}\mathcal{D}(\sigma_z)$, while the cavity mode experiences decay $\mathcal{D}(a)$.
The qubit is assigned the same decay and dephasing rates as those used in the Heisenberg model.
The qubit-cavity coupling strength is set to $g=2\,\mathrm{MHz}$, and the cavity mode is taken to have a photon decay time of $T_r=4\,\mu\mathrm{s}$; these parameters are chosen to be consistent with Ref.~\cite{LangfordNK2017DQRS}.
This relatively weak coupling strength is required to enable accurate digital quantum simulation of the Rabi model using Trotterisation~\cite{LangfordNK2017DQRS}.
For the RMT analyses, the cavity mode is truncated to a maximum of $\text{dim}_c=4$ photons, consistent with the analysis in Ref.~\cite{kargi2025quantum}, which is sufficient to provide enough eigenvector components for a meaningful comparison with RMT predictions.
For dynamical simulations, however, the cavity mode is truncated to a maximum of $\text{dim}_c=20$ photons to ensure that the dynamics are accurately captured, as the photon number can increase during the evolution.

\section{Persistence of the Trotterisation Threshold in Fidelity Dynamics}\label{Supp:FidelityDynamics}


\begin{figure*}[!t]
    \centering
    \includegraphics{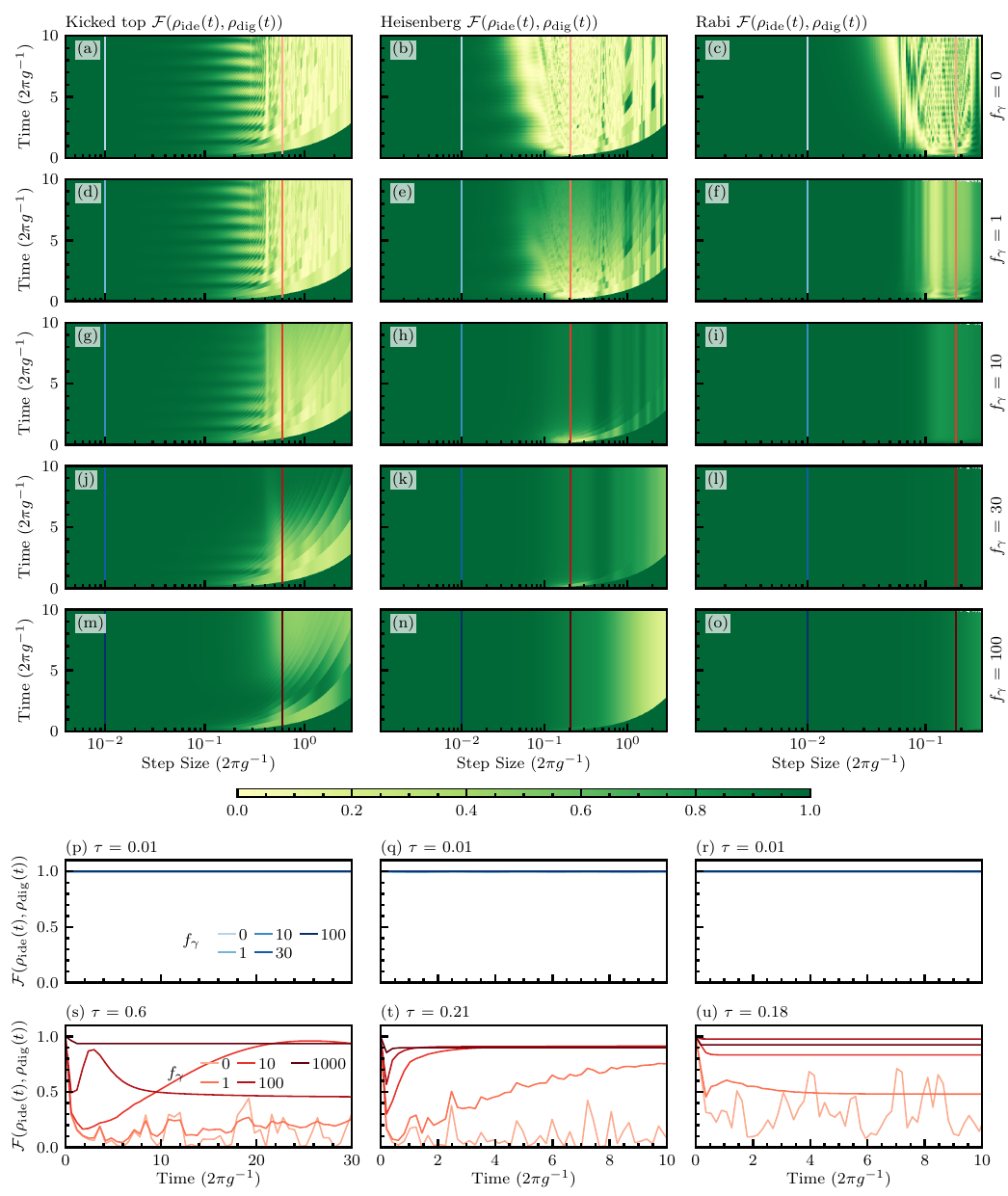}   
    \caption[Fidelity]{Simulation fidelity $\mathcal{F}(\rho_{\text{ide}}(t), \rho_{\text{dig}}(t))$ of the Trotterised evolution with respect to the target open-system evolution for the kicked top, Heisenberg, and Rabi models, corresponding to the first, second, and third columns, respectively.   
    The parameters are the same as in Fig.~\ref{fig:Dynamics}.
    The first five rows show the time evolution of the fidelity under open-system DQS at different dissipation strength scaling factors $f_\gamma = 0.01, 1, 10, 30, 100$ respectively, plotted as a function of the Trotter step size $\tau$ (x-axes) and simulation time (y-axes).
    (p-r) Pre-threshold and (s-u) post-threshold time traces of the fidelity for different dissipation strengths.
    }
    \label{fig:Fidelity}

\end{figure*}


\begin{figure*}[!t]
    \centering

    \includegraphics{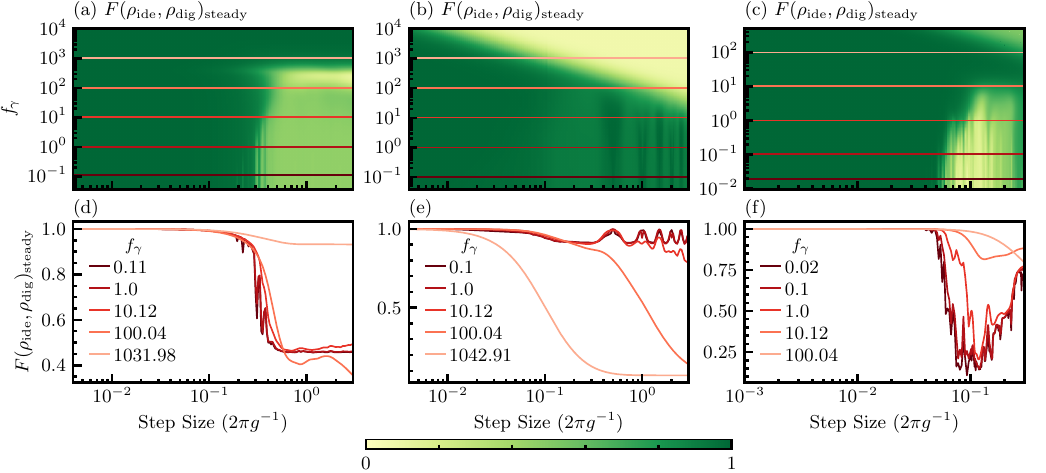}

    \caption[Steady-state fidelity]{
    Steady-state values of simulation fidelity $\mathcal{F}(\rho_{\mathrm{ide}},\rho_{\mathrm{dig}})_\mathrm{steady}$ in open-system DQS for the kicked top ($j=10$), Heisenberg (chain of $N=5$ qubits), and Rabi (cavity truncation dimension $\text{dim}_c=20$) models, corresponding to the first, second, and third columns. 
    Panels (a--c) show the fidelity as a function of $\tau$ and $f_\gamma$, while (d--f) show traces at selected values of $f_\gamma$.}
    \label{fig:SteadyCombined2}
\end{figure*}


The fidelity $\mathcal{F}(\rho_{\text{ide}}(t), \rho_{\text{dig}}(t))$ of the Trotterised evolution with respect to the target open-system evolution can be used to quantify the accuracy of the digital quantum simulation.
The fidelity is defined as $\mathcal{F}(\rho_{\text{ide}}(t), \rho_{\text{dig}}(t)) = \left(\text{Tr}\left[\sqrt{\sqrt{\rho_{\text{ide}}(t)}\rho_{\text{dig}}(t)\sqrt{\rho_{\text{ide}}(t)}}\right]\right)^2$, where $\rho_{\text{ide}}(t)$ and $\rho_{\text{dig}}(t)$ are the density matrices of the ideal and digital evolutions, respectively.
The Trotterisation threshold can be identified by analysing the time evolution of the fidelity for different Trotter step sizes and dissipation strengths.

Figure~\ref{fig:Fidelity} shows the time evolution of the fidelity for the kicked top, Heisenberg, and Rabi models in the first, second, and third columns, respectively. 
The first five rows show the time evolution of the fidelity under open-system DQS at different dissipation strength scaling factors $f_\gamma = 0.01, 1, 10, 30, 100$ respectively, plotted as a function of the Trotter step size $\tau$ (x-axes) and simulation time (y-axes).
The last two rows show the pre-threshold and post-threshold time traces of the fidelity for different dissipation strengths.
The initial state of the kicked top model is a maximal spin state, the initial state of the Heisenberg model has only the first qubit in the chain excited, and the initial state of the Rabi model has an excited qubit, with the cavity mode in the vacuum.

A clear threshold can be observed in the time evolution of the fidelity for all three models and for dissipation strengths less than $f_\gamma \lesssim 30$ [Fig.~\ref{fig:Fidelity}~(a-k)], which is consistent with the Trotterisation threshold identified in the time-evolution of local observables. 
In pre-threshold regimes, the fidelity remains close to $1$ for all dissipation strengths, indicating that the Trotterised evolution accurately approximates the target open-system evolution [Fig.~\ref{fig:Fidelity}~(p-r)], and beyond the threshold, the fidelity decays rapidly.
However, as the dissipation rate increases, the fidelity beyond the threshold improves [Fig.~\ref{fig:Fidelity}~(s-u)] because the dissipative processes begin to dominate the dynamics, and the Trotterisation errors become less significant in comparison.

The persistence of the Trotterisation threshold in the fidelity dynamics suggests that the threshold may be present even in the steady-state regime. 
Figure~\ref{fig:SteadyCombined2} shows the steady-state fidelity $\mathcal{F}(\rho_{\text{ide}}, \rho_{\text{dig}})_\text{steady}$ of the Trotterised evolution with respect to the target open-system evolution for the kicked top, Heisenberg, and Rabi models in the first, second, and third columns, respectively. 
Figures~\ref{fig:SteadyCombined2}~(a-c) show the steady-state fidelity as a function of the Trotter step size $\tau$ (x-axes) and dissipation rate scaling factors $f_\gamma$ (y-axes), and Figs.~\ref{fig:SteadyCombined2}~(d-f) show traces of the steady-state fidelity at different dissipation rate scaling factors $f_\gamma$. 
The figure shows a clear threshold in the steady-state fidelity for all three models, which is consistent with the Trotterisation threshold identified in the time-evolution of local observables and the time evolution of the fidelity. 
The threshold vanishes as the dissipation rate increases, illustrating that the dissipative processes begin to dominate the coherent part of the dynamics completely. 


%

\end{document}